\documentclass[journal = nalefd]{achemso}
\usepackage{xr-hyper, hyperref, graphicx, amsmath, amssymb, xcolor, bm}
\graphicspath{{figures/}}

\author{Valentin A. Semkin}
\affiliation{Center for Photonics and 2D Materials, Moscow Institute of Physics and Technology, Dolgoprudny 141700, Russia}
\email{semkin.va@phystech.edu}
\author{Aleksandr V. Shabanov}
\affiliation{Center for Photonics and 2D Materials, Moscow Institute of Physics and Technology, Dolgoprudny 141700, Russia}
\author{Dmitry A. Mylnikov}
\affiliation{Center for Photonics and 2D Materials, Moscow Institute of Physics and Technology, Dolgoprudny 141700, Russia}
\author{Mikhail A. Kashchenko}
\affiliation{Center for Photonics and 2D Materials, Moscow Institute of Physics and Technology, Dolgoprudny 141700, Russia}
\alsoaffiliation{Programmable Functional Materials Lab, Brain and Consciousness Research Center, Moscow, 121205, Russia}
\author{Ivan K. Domaratskiy}
\affiliation{Center for Photonics and 2D Materials, Moscow Institute of Physics and Technology, Dolgoprudny 141700, Russia}
\author{Sergey S. Zhukov}
\affiliation{Center for Photonics and 2D Materials, Moscow Institute of Physics and Technology, Dolgoprudny 141700, Russia}
\author{Dmitry A. Svintsov}
\affiliation{Center for Photonics and 2D Materials, Moscow Institute of Physics and Technology, Dolgoprudny 141700, Russia}

\title{Zero-bias photodetection in 2d materials via geometric design of contacts }

\begin{document}
\maketitle

\begin{abstract}
Structural or crystal asymmetry are necessary conditions for emergence of zero-bias photocurrent in light detectors. Structural asymmetry has been typically achieved via $p-n$ doping being a technologically complex process. Here, we propose an alternative approach to achieve zero-bias photocurrent in 2d material flakes exploiting the geometrical non-equivalence of source and drain contacts. As a prototypical example, we equip a square-shaped flake of PdSe$_2$ with mutually orthogonal metal leads. Upon uniform illumination with linearly-polarized light, the device demonstrates non-zero photocurrent which flips its sign upon 90$^\circ$ polarization rotation. The origin of zero-bias photocurrent lies in polarization-dependent lightning-rod effect. It enhances the electromagnetic field at one contact from the orthogonal pair, and selectively activates the internal photoeffect at the respective metal-PdSe$_2$ Schottky junction. The proposed technology of contact engineering can be extended to arbitrary 2d materials and detection of both polarized and natural light.
\end{abstract}


A photodetector should possess some asymmetry or 'preferred direction' to generate non-zero photocurrent. This stems from the fact that radiation intensity is a scalar, while photocurrent is a vectorial quantity. In the simplest case of photoconductors and bolometers, direct bias current sets such direction. However, this approach suffers from current-induced noise. The bias-free detection, enabling low noise, requires the asymmetry of material crystal lattice or the structure of detector itself. In the latter case, being the most practical, the asymmetric doping of the detector body ($p-n$ junction) provides the necessary asymmetry.

Two-dimensional materials (2DMs) are expected to enrich the functionalities of light detectors by providing short response times~\cite{Ultrafast_graphene,Ultrafast_vdW}, compatibility with silicon waveguides~\cite{Chip-integrated-graphene} due to atomically thin body, and in-situ tunability of optical properties~\cite{Kono-tunable-properties}. Implementation of conventional $p-n$ junctions with 2d materials, however, faces technological challenges. Indeed, the technologies of chemical doping for 2DMs are not yet established and inevitably lead to mobility degradation~\cite{Liu2014,Graphene_Lateral_PNJ}. Field-effect doping of 2DMs~\cite{Castilla2019,Mylnikov2023} resolves the mobility issue, but may crop the response times due to extra gate capacitance. Thus, the asymmetry in 2d photodetectors, different from conventional $p-n$ junction, is of high demand.

It has been already noted that a large fraction of photocurrent in 2DM-based detectors is formed at the Schottky junction between the 2d body and metal contacts~\cite{Mueller2010,Echtermeyer2014,Tielrooij2015,Cai2014,Bandurin2018}. This occurs both for visible light, where the built-in field separates electron-hole pairs~\cite{Mueller2010,Echtermeyer2014}, and for long-wavelength radiation where thermoelectric effects and current-voltage nonlinearities take place~\cite{Cai2014,Bandurin2018}. In a typical detector structure with 2DM placed between identical metals, the photocurrents of two Schottky junctions are exactly compensated, in agreement with symmetry requirements discussed above. Numerous works in the field dealt with contacts of dissimilar metals~\cite{Mueller2010,Cai2014,Gayduchenko2018,Semkin}, which, however, requires extra technological steps and careful selection of work functions. If it was possible to 'shadow' one of the Schottky junctions, the necessary asymmetry would be introduced, and one would get the desired zero-bias low-noise photodetector based on a 2d material.

Here, we show that geometric inequivalence of detector contacts results in the desired 'shadowing' of one of the Schottky junctions in the near-field, and thus in finite photocurrent at zero bias and uniform illumination. We may suggest similar effects took place in the so-called 'geometric rectifiers' with non-uniform channel~\cite{Zhou_source-drain_width,Auton2017,Geometric_rectifiers,Chen_Nonuniform_channel}, but this possibility was not realized. Alternative attempts to modify the spatial distribution of local fields exploited nano-antennas~\cite{Wei2020,Wei2021} and gratings~\cite{Popov_2015,Ganichev_Grating} placed above the device, which is also challenging technologically. Here, as a practical consequence, we achieve zero-bias polarization-sensitive detection in technologically simple two-terminal structures. The idea is generic and can be applied to arbitrary 2d materials. 

\begin{figure*}
    \includegraphics{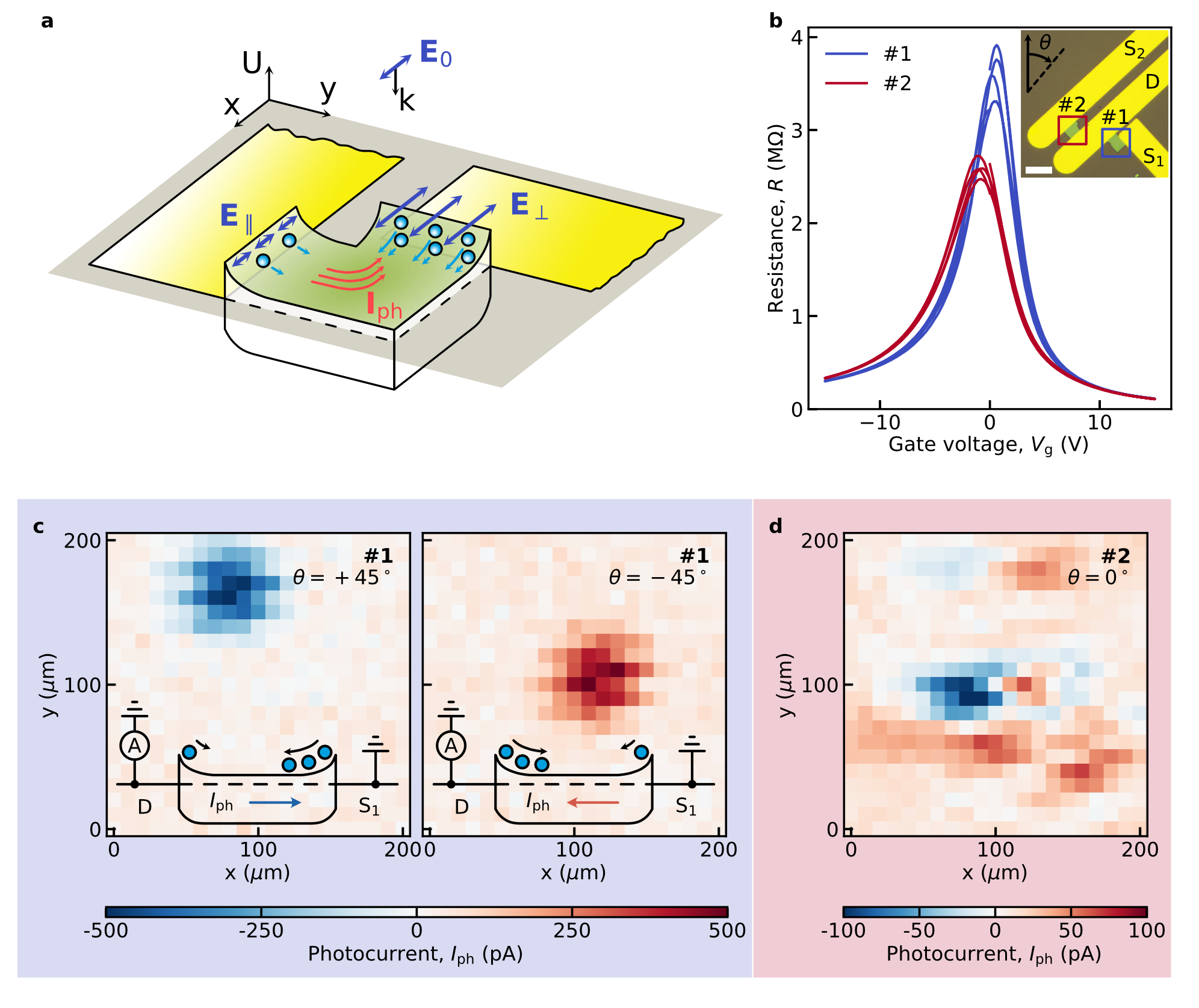}
    \caption{\label{fig1}{\bf Light detection with asymmetric corner-type and symmetric slit-type detectors.} {\bf a,} Illustration of the operating principle for the corner-type detector: incident linearly-polarized field is enhanced by the metal contact orthogonal to ${\bf E}$-field, and suppressed by another contact. The Schottky junction at the 'active' contact generates large photocurrent, which cannot be compensated by small current at the passive junction. {\bf b,} Gate dependencies of the devices' resistance. Inset: micro-photograph of the devices and the reference direction for polarization angles $\theta$. Scale bar is 10 $\mu$m.  {\bf c,d,} Spatial maps of photocurrent recorded at $\lambda_0~=~8.6~\mu$m and zero bias voltage. {\bf c,} Map for corner-type detector \#1 at a gate voltage $V_{\rm g}~=~20~$V and polarization angles $\theta~=~+45^\circ$ (left) and $\theta~=~-45^\circ$ (right). The shift of the photoresponse spot is caused by the polarizer rotation. Insets: band diagrams with device connection scheme and photocurrent directions. {\bf d,} Map for reference detector \#2 at a gate voltage $V_{\rm g}~=~10$ V and a polarization angle $\theta~=~0^\circ$.}
\end{figure*}

For prototypical example, we fabricate and study a detector with rectangular channel and metal contacts to its orthogonal sides. Its operating principle is illustrated in Fig.~\ref{fig1}a and its micro-photograph is shown in the inset of Fig.\ref{fig1}b, \#1.
The linearly-polarized radiation with ${\bf E}$-field orthogonal to one of metal contacts is enhanced locally via to dynamic lightning-rod effect~\cite{Nikulin2021,Semkin}. Near another contact, the incident field is suppressed due to dynamic screening currents in metal. As a result, strong photocurrent at the 'electromagnetically active' Schottky junction is no more compensated by the current at the passive one. Implementation of this idea requires the region of near-field enhancement ($\sim \lambda_0/100...\lambda_0/10$) to overlap with metal-2DM Schottky junction of typical length $l_{\rm J} \approx 100$ nm. Here, we use the mid-infrared illumination with wavelength $\lambda_0 = 8.6$ $\mu$m, for which this condition is best fulfilled. 

The channel of our corner-type detector is made of quasi two-dimensional PdSe$_2$ [Fig.~\ref{figS1}, \#1], which already demonstrated strong photoresponse at mid-infrared range~\cite{Long2019}. The channel thickness $t$ is $\sim 13$~nm [Fig.~\ref{figS2}]. The channel size is $4.5\times6~\mu$m$^2$, which is below the wavelength of incident light. A symmetric reference detector \#2 was fabricated with the same optical thickness and a comparable channel size $5\times3.5$ $\mu$m$^2$. The symmetric device has contacts to opposite rectangle sides. Both devices are deposited on a silicon substrate acting as a back gate, covered with 300 nm SiO$_2$ gate dielectric. The devices share the same drain terminal D used for measurements of current and different grounded sources S$_1$ and S$_2$. Two devices have very similar dependencies of resistance $R$ on gate voltage $V_{\rm g}$, as shown in Fig.~\ref{fig1}b.

To prove the suggested operating principle of contact-engineered corner-type device, we present its spatially-resolved photocurrent maps. They were recorded by illuminating the device with linearly polarized, focused beam from IR quantum cascade laser with wavelength $\lambda_0~=~8.6~\mu$m and power $P~\approx~20$ mW. Once the radiation polarization is orthogonal to the source S$_1$, one observes a single bright photocurrent spot on the map, as shown in Fig.~\ref{fig1}c, left panel. The situation is in stark contrast to our symmetric slit-type detector \#2 [Fig.~\ref{fig1}d], and to many previous observations~\cite{Lemme,Badioli2014}. There, two nearby photocurrent spots of different sign and reduced magnitude are observed.

Two weak photocurrent spots in nearly-symmetric detectors represent a result of competing current generation events at the source and drain junctions. The light spot covers both terminals in sub-wavelength symmetric detectors, and local intensities at the terminals differ by a small amount, depending on beam position. Once the beam is shifted to the top position in Fig.~\ref{fig1}d, the source junction gains slightly larger local intensity, and its photocurrent prevails over the drain. The situation is reverted when shifting the spot downwards to the drain. 

We do not observe any traces for secondary photocurrent spot in corner-type device, Fig.~\ref{fig1}c, left panel. It implies that one of its contacts (hereby source) generates dominant current independent of the beam position. The maximum photocurrent in corner-type device is $\sim 0.5~$nA, which 5x times exceeds the photocurrent at each of two spots for the symmetric slit-type device. Rotation of light polarization by 90$^\circ$ results in inversion of photocurrent sign in device \#1, as shown in Fig.~\ref{fig1}c, right panel. The inversion is explained by the activation of an opposite Schottky junction at the drain via polarization-dependent lightning-rod effect. 

Further confirmations of the suggested operating principle are obtained by recording the dependencies of photocurrent $I_{\rm ph}$ on bias voltage $V_{\rm sd}$, gate voltage $V_{\rm g}$, and polarization angle $\theta$ counted off from the diagonal of square-shaped channel [Fig.~\ref{figS3}, \ref{figS4}]. These characteristics are conveniently analyzed by polynominal fits of $I_{\rm ph}$ vs $V_{\rm sd}$ [Fig.~\ref{figS5}]:
\begin{equation}
I_{\rm ph} = I_0 + G_{\rm ph} V_{\rm sd} + \frac{1}{2}\frac{d^2 I_{\rm ph}}{dV^2_{\rm sd}}V^2_{\rm sd} + \frac{1}{6}\frac{d^3 I_{\rm ph}}{dV^3_{\rm sd}}V^3_{\rm sd},
\label{eq:approx}
\end{equation}
where $I_0$ is the zero-bias photocurrent, $G_{\rm ph}$ is the photoconductivity, and higher-order terms up to the cubic describe channel nonlinearities.

\begin{figure*}[ht]
    \includegraphics{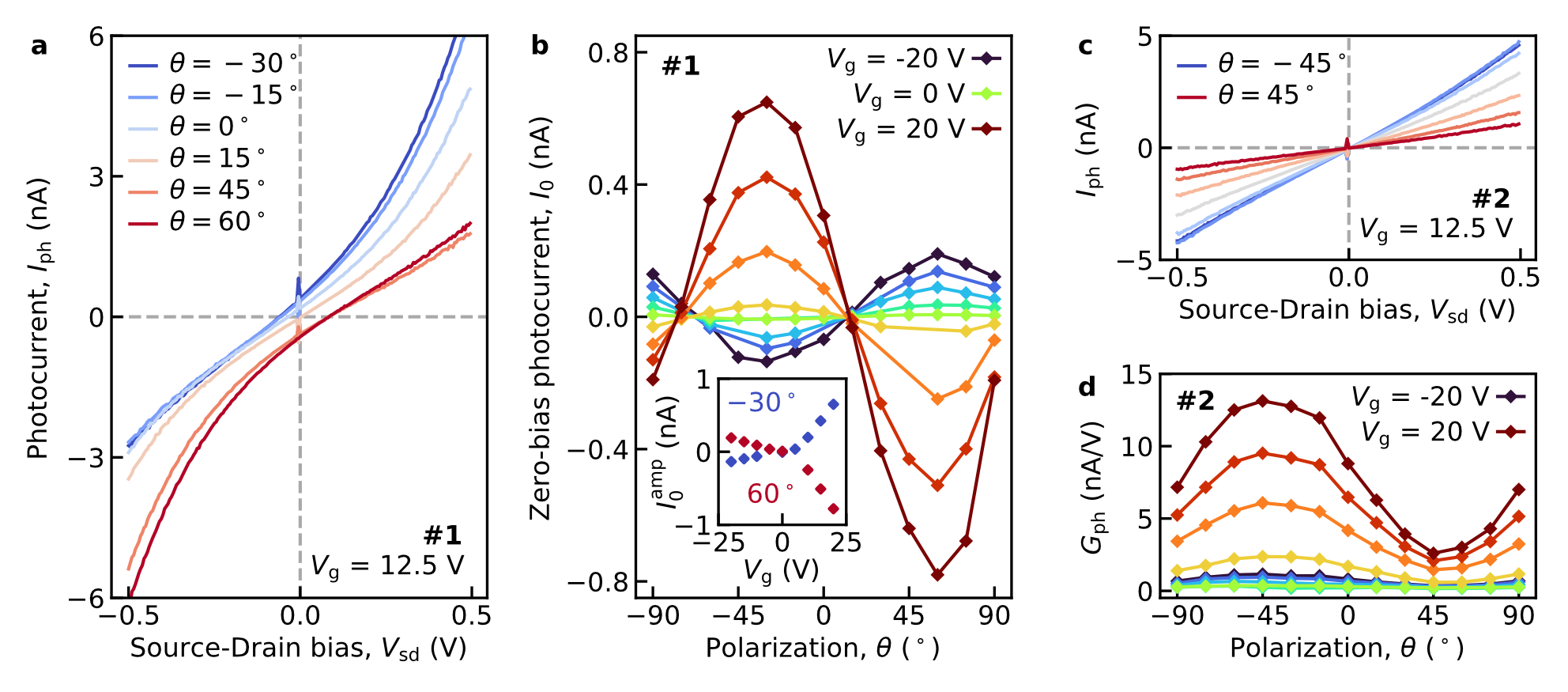}
    \caption{\label{fig2}{\bf Mid-infrared (8.6 $\bm{\mu}$m) polarization-sensitive photoresponse.} {\bf a,c,} A series of photocurrent $I_{\rm ph}$ dependencies on the bias voltage $V_{\rm sd}$ at different polarization angles $\theta$ for corner-type detector $\#1$ ({\bf a}) and for reference detector $\#2$ ({\bf c}). {\bf b,} Extracted series of zero-bias photocurrents $I_0$ for device $\#1$ vs polarization angle at different gate voltages $V_{\rm g}$ with step $\Delta V_{\rm g}~=~5~$V. Inset: Amplitude of zero-bias photocurrent $I_0^{\rm amp}$ vs $V_{\rm g}$ at $\theta~=~-30^{\circ}$ and $\theta~=~+60^{\circ}$. {\bf d,} Extracted series of photoconductivity $G_{\rm ph}$ vs polarization angle for device $\#2$ at different $V_{\rm g}$ taken with step $\Delta V_{\rm g}~=~5~$V. See Fig.~\ref{figS6} for $I_0(\theta, V_{\rm g})$ series for \#2 and $G_{\rm ph}(\theta, V_{\rm g})$ series for \#1.}
\end{figure*}

As shown in Fig.~\ref{fig2}a, the corner-type device indeed demonstrates a notable zero-bias photocurrent $I_0$. Its absolute value is maximized and has opposite signs for $\theta = -30^\circ$ and $\theta = 60^\circ$. These angles correspond approximately to ${\bf E}$-field orthogonal to the drain and source contacts, respectively. Interestingly, the non-linearity of photocurrent is polarization-sensitive, and manifests itself for $\theta>0$ at $V_{\rm sd} < 0$ or for $\theta < 0$ for $V_{\rm sd} > 0$.

Application of gate voltage considerably modifies the zero-bias current in corner-type device. As seen from Fig.~\ref{fig2}b, larger absolute values of $|V_{\rm g}|$ result in larger photocurrents, while sign flip of $V_{\rm g}$ leads to reversal of $I_0$. The first fact is readily explained by reduced PdSe$_2$ bulk resistance at high carrier density, controlled by $V_{\rm g}$ [inset of Fig.~\ref{fig2}b]. The polarity of $I_0$ is linked to the sign of majority carriers. When the D-terminal of corner-type device is active ($\theta \approx -30^\circ$) and the channel is $n$-doped ($V_{\rm g} > 0$), both photocurrent and photovoltage are positive. It implies the motion of photogenerated electrons away from the drain. Such motion can occur upon photovoltaic effect in a band profile shown in the inset of Fig.~\ref{fig1}c. The observed sign of zero-bias current can be alternatively explained by thermal diffusion of majority carriers from 'hot' contact to the 'cold' bulk. Discrimination between photovoltaic and thermoelectric scenarios requires more careful determination of PdSe$_2$ band gap~\cite{Nishiyama}.

Similar measurements and data processing were performed for the symmetric slit-type detector \#2. The dependencies of photocurrent on bias voltage and polarization angle hereby show a completely different qualitative behavior [Fig.~\ref{fig2}c]. The zero-bias photocurrent is absent (least for centered illumination) and the dependencies $I_{\rm ph}(V_{\rm sd})$ are almost linear. The photoconductivity $G_{\rm ph}$ of the symmetric device, shown in Fig.~\ref{fig2}d, displays strong, approximately three-fold, variation with polarization angle. It is maximized for ${\bf E}$-field perpendicular to metal contacts, and drops to minimum upon $90^\circ$ polarization rotation. Again, these observations are explained by lightning-rod effect at the metal contacts. Indeed, the photoconductivity is maximized at maximum local field in the channel, which occurs precisely for ${\bf E}$-field orthogonal to the contacts. The sign of photoconductivity in our measurements is positive, which hints on the possibility of interband electron-hole generation.

We complete our discussion of contact-engineered detectors by presenting the simulation results for local electromagnetic fields. This enables us not only to prove the suggested physics of polarization-dependent lightning-rod effect, but also to indicate possible optimization directions (see Methods for technical details).

\begin{figure*}[ht]
    \includegraphics{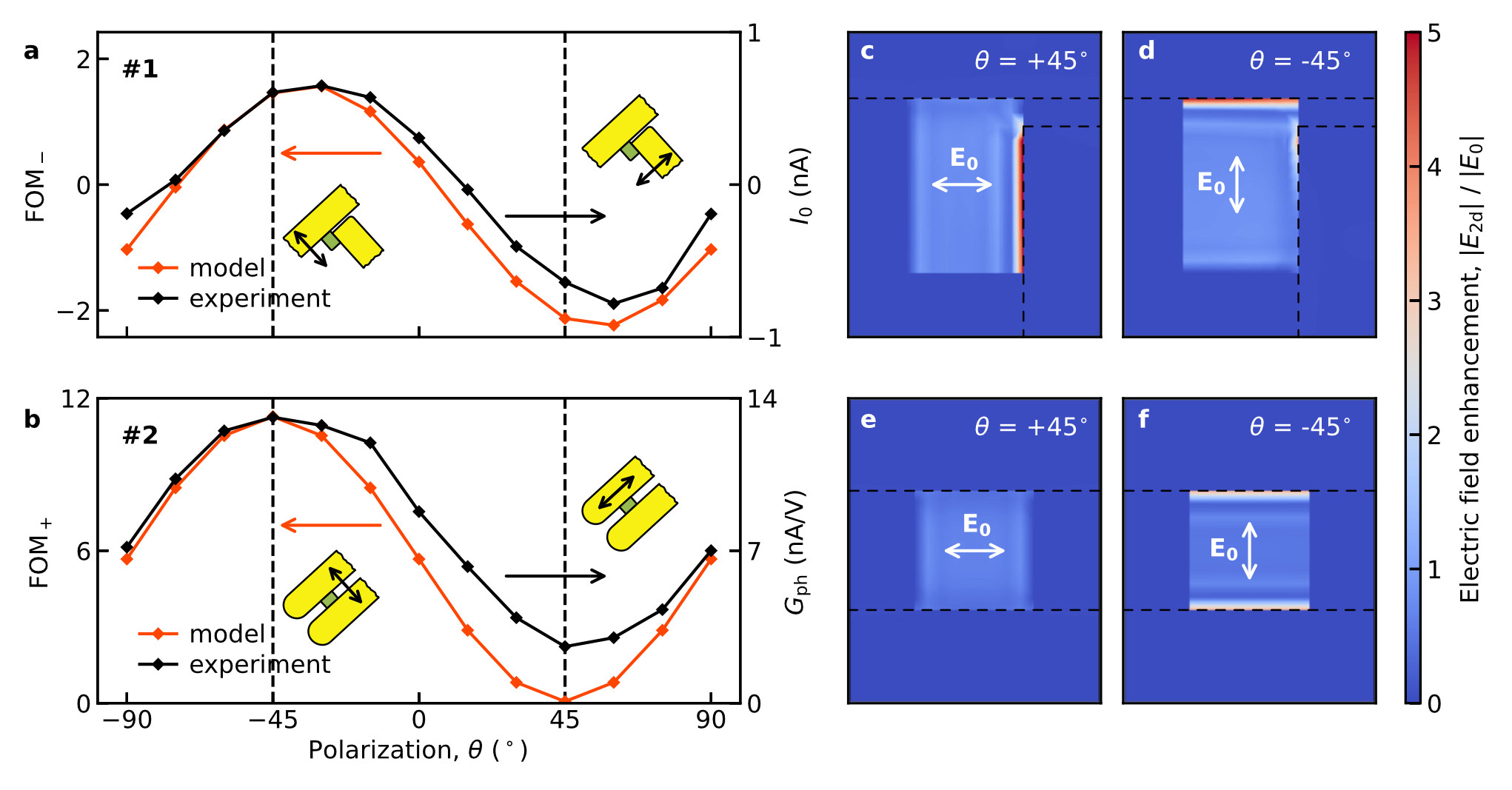}
    \caption{\label{fig3}{\bf Simulation of contact-engineered detectors.} {\bf a,} Comparison of the measured zero-bias photocurrent $I_0(\theta)$ with the simulation of the normalized difference of average local intensities at source and drain FOM$_-$ for device \#1. {\bf b,} Comparison of the measured photoconductivity $G_{\rm ph}(\theta)$ with the simulation of normalized sum of average local intensities at source and drain FOM$_+$ for the device \#2. Both experimental curves in {\bf a} and {\bf b} were measured at gate voltage $V_{\rm g}$ = 20 V. Insets: polarization of the incident light at $\theta~=~\pm45^{\circ}$ is shown by the black arrows relative to the device schematics. {\bf c-f,} Simulated local electric field in the 2d material plane normalized by the incident field, $|{\bf E}_{\rm 2d}/{\bf E}_0|$ for the corner-type ({\bf c} and {\bf d}) and reference slit-type ({\bf e} and {\bf f}) detectors. Polarization of the incident light ${\bf E}_0$ is shown with the white arrow.}
\end{figure*}

The simulation of local field distribution under plane-wave linearly polarized illumination [Fig.~\ref{fig3}c,d] confirms strong field enhancement near one metal pad and its suppression near another. Approaching the 'active pad', the field diverges as inverse square root of distance. At shortest distances, it saturates to finite value inversely proportional to 2d material conductivity. An interesting feature seen in simulations is oscillation of local field intensity along the channel indicating that resonant effects in PdSe$_2$ flake may affect the photoresponse.

A perfect asymmetrically engineered detector should have very small local field at one contact, and large field at another one. It is also natural to require local field enhancement in a well-designed device, compared to the incident field $E_0$. Based on these arguments, we choose the electromagnetic figure of merit  for asymmetric device (FOM$_-$) as the difference of average local intensities at source and drain, $E_{\rm S}^2$ and $E_{\rm D}^2$, normalized by the incident field intensity $E_0^2$:
\begin{equation}
\label{Eq-FOM}
{\rm FOM}_- = \frac{\langle E_{\rm S}^2\rangle - \langle E_{\rm D}^2\rangle}{E_0^2}.
\end{equation}
Averaging is performed along the length of contact, while the distance between metal pad and the point where field is calculated is taken to be 100 nm, an approximate width of the Schottky junction~\cite{Lee_contacts_and_edges}.

We expect that FOM$_-$ given above, Eq.~(\ref{Eq-FOM}), serves as a measure of zero-bias current independent of particular detection physics. Particularly, it measures the difference between numbers of electron-hole pairs generated at two Schottky junctions in the case of internal photovoltaic effect~\cite{Afzal_PdSe2_photovoltaic}. It is also proportional to the difference in electron temperatures at the two junctions, which is proportional to photocurrent if photothermoelectric effect takes place~\cite{Dai}. In any case, the proportionality between FOM$_-$ and photocurrent $I_0$ indeed experimentally holds. We show it in Fig.~\ref{fig3}a by comparing the computed FOM$_-(\theta)$ (orange line) and the measured zero-bias photocurrent (black line). These quantities coincide well up to a dimensional multiplier. The simulation captures well the tiny details of experiment. Particularly, the maxima of FOM$_-$ and photocurrent appear not exactly for ${\bf E}$-field orthogonal to the contacts, but are shifted by $\sim 15^\circ$ from the expected position. This feature stems from deviations of detector structure from perfectly symmetric square PdSe$_2$ channel [Fig.~\ref{figS7}a].

The polarization dependence of photocurrent in a symmetric device is governed by the same vectorial lightning-rod effect, though the signal emerges only at finite bias. The photoconductivity $G_{\rm ph}$ of the symmetric device should be proportional to the absorbed power at the Schottky junctions, and, in turn, to the quantity ${\rm FOM}_+ = (\langle E_{\rm S}^2\rangle + \langle E_{\rm D}^2\rangle)/E_0^2.$ As shown in Fig.~\ref{fig3}b, the simulated FOM$_+(\theta)$ matches well the measured photoconductivity. The agreement becomes even better if we compute the electromagnetic power absorbed in the whole PdSe$_2$ channel, not only at its contacts [Fig.~\ref{figS7}b].

In conclusion, we have shown that geometric engineering of contacts to 2d materials provides a unique opportunity to achieve zero-bias photodetection with low noise. The technique applies to arbitrary 2d materials and is readily scalable to arbitrary wavelengths. The only requirement for achievement of appreciable photosignal is the presence of rectifying Schottky junction at metal-2d semiconductor contact, which is the case unless the ohmic contact is fabricated intentionally~\cite{Contact_engineering}. The physics of photocurrent generation at the junction, again, can be arbitrary. It may include but not limited to photovoltaic, photothermoelectric effects and direct rectification by junction $I(V)$-nonlinearities.

It is important that the demonstrated pair of photodetectors is able to resolve the linear polarization state of the incident light~\cite{Wei2020}. Namely, by normalizing the signal of corner-type detector to the signal of slit-type one, one gets rid of photocurrent dependence on light power. The dependence of normalized current on $\theta$ can be recorded once and further used for unambiguous determination of polarization angle within the $90^\circ$-range. Already the sign of photocurrent in corner-type device tells us unambiguously which quadrant the polarization angle belongs to. This property can be important for optical communications using two independent orthogonal polarization channels for increased bit capacitance.

An arbitrary geometric inequivalence of two contacts to 2d channel should result in finite zero-bias photocurrent~\cite{Zhou_source-drain_width,Chen_Nonuniform_channel}. To make this photocurrent large, the inequivalence should be strong. We have shown that it can be quantitatively characterized by the figure of merit (\ref{Eq-FOM}) being the ratio of local intensity difference at the two contacts, normalized by intensity of the incident light. In the proposed design with orthogonal source and drain contacts, the numerical value of FOM$_-$ is moderate and reaches $\sim 2$ in absolute value. Yet, there are no fundamental limitations to this quantity. According to our simulations, further seven-fold enhancement of FOM$_-$ is possible upon optimization of contacts' width and SiO$_2$ thickness (see Supporting Information, Section IV).

We have shown that a versatile tool to achieve electromagnetically inequivalent contacts lies in exploiting polarization-dependent field enhancement by keen metal edges~\cite{Nikulin2021,Semkin}. Such effect may be even more pronounced for contacts with in-plane patterning, which may have either regular (e.g. sawtooth) or fractal~\cite{Polini_fractal} shape. Another option toward electromagnetic asymmetry may rely on optical resonance in one metal pad and anti-resonance in another. Optimized shapes of metal contacts can be accessed with machine-learning techniques~\cite{Campbell_MLDesign}. To conclude, the technique of geometrically dissimilar contacts to 2d materials open wide opportunities in low-noise zero-bias detection with possible option of polarization resolution.

{\bf Methods.} PdSe$_2$ flakes were prepared by micro-mechanical cleavage of the bulk crystal and deposited on lightly doped SiO$_2$/Si substrate [Fig.~\ref{figS1}]. Their thickness $t\sim 13$ nm was verified using atomic force microscopy [Fig.~\ref{figS2}]. Top contacts to PdSe$_2$ were achieved by magnetron sputtering of Ti/Au (3/60 nm).

Optical measurements were performed in evacuated chamber with residual gas pressure $P\sim 10^{-4}$ Torr. The radiation was fed from quantum cascade laser (QCL) with central wavelength $\lambda_0=8.6$ $\mu$m. A $\lambda_0/4$-waveplate with axes rotated by $45^\circ$ to initial polarization direction of laser radiation and a polarizer were introduced between laser and sample. This enabled power-preserving polarization rotation in our experiment. We used internal modulation of QCL current with $f_{\rm mod} = 911$ Hz, which enabled lock-in measurements of the photocurrent.

Simulations were performed in a CST Microwave studio package using finite-element method at a single frequency. Metal contacts and PdSe$_2$ flake were treated as infinitely thin (sheet) conductors with areal impedances $Z_{\rm m} = 0$ and $Z_{\rm PdSe2} = 10^{-4}Z_0$, respectively, where $Z_0=377$ Ohm is the free-space impedance. The definite value of PdSe$_2$ impedance does not affect the shape of FOM$_{\pm} (\theta)$ dependence as soon as $Z_{\rm PdSe_2} \ll Z_0$. The latter fact simply implies that absorbance by extended PdSe$_2$ film is much less than unity, which is an experimentally established fact~\cite{Ermolaev2021}. Wavelength-dependent dielectric constants SiO$_2$ for simulations were used from Ref.~\citenum{Kischkat}.

\section*{Data Availability Statement}
The data that support the findings of this study are available from the corresponding author upon reasonable request.

\section*{Supporting Information}
(I) Devices fabrication and characterization, (II) Photocurrent measurement details, (III) Simulation details, (IV) Optimization of corner-type detector (PDF)

\section*{Acknowledgments}
This work was supported by the Russian Science Foundation (Grant No. 21-72-00078). The devices were fabricated using the equipment of the Center of Shared Research Facilities (Moscow Institute of Physics and Technology). We are grateful to V.V. Dremov for experimental assistance.

\section*{Author Contributions}
D.S. conceived and supervised the project. V.S and D.M performed transport and photoresponse measurements. V.S and D.S analyzed experimental data. A.S. and D.S. developed theoretical model and performed simulations. M.K and I.D. fabricated devices. S.Z. assisted in tuning of optical setups. V.S. and D.S. wrote the manuscript, with input from all the co-authors. All authors contributed to the discussions.

\section*{Competing interests}
The authors declare no competing financial interest.

\bibliography{references}

\providecommand{\latin}[1]{#1}
\makeatletter
\providecommand{\doi}
  {\begingroup\let\do\@makeother\dospecials
  \catcode`\{=1 \catcode`\}=2 \doi@aux}
\providecommand{\doi@aux}[1]{\endgroup\texttt{#1}}
\makeatother
\providecommand*\mcitethebibliography{\thebibliography}
\csname @ifundefined\endcsname{endmcitethebibliography}
  {\let\endmcitethebibliography\endthebibliography}{}
\begin{mcitethebibliography}{37}
\providecommand*\natexlab[1]{#1}
\providecommand*\mciteSetBstSublistMode[1]{}
\providecommand*\mciteSetBstMaxWidthForm[2]{}
\providecommand*\mciteBstWouldAddEndPuncttrue
  {\def\EndOfBibitem{\unskip.}}
\providecommand*\mciteBstWouldAddEndPunctfalse
  {\let\EndOfBibitem\relax}
\providecommand*\mciteSetBstMidEndSepPunct[3]{}
\providecommand*\mciteSetBstSublistLabelBeginEnd[3]{}
\providecommand*\EndOfBibitem{}
\mciteSetBstSublistMode{f}
\mciteSetBstMaxWidthForm{subitem}{(\alph{mcitesubitemcount})}
\mciteSetBstSublistLabelBeginEnd
  {\mcitemaxwidthsubitemform\space}
  {\relax}
  {\relax}

\bibitem[Xia \latin{et~al.}(2009)Xia, Mueller, Lin, Valdes-Garcia, and
  Avouris]{Ultrafast_graphene}
Xia,~F.; Mueller,~T.; Lin,~Y.~M.; Valdes-Garcia,~A.; Avouris,~P. {Ultrafast
  graphene photodetector}. \emph{Nature Nanotechnology} \textbf{2009},
  \emph{4}, 839--843\relax
\mciteBstWouldAddEndPuncttrue
\mciteSetBstMidEndSepPunct{\mcitedefaultmidpunct}
{\mcitedefaultendpunct}{\mcitedefaultseppunct}\relax
\EndOfBibitem
\bibitem[Massicotte \latin{et~al.}(2016)Massicotte, Schmidt, Vialla,
  Sch{\"{a}}dler, Reserbat-Plantey, Watanabe, Taniguchi, Tielrooij, and
  Koppens]{Ultrafast_vdW}
Massicotte,~M.; Schmidt,~P.; Vialla,~F.; Sch{\"{a}}dler,~K.~G.;
  Reserbat-Plantey,~A.; Watanabe,~K.; Taniguchi,~T.; Tielrooij,~K.~J.;
  Koppens,~F. H.~L. {Picosecond photoresponse in van der Waals
  heterostructures}. \emph{Nature Nanotechnology} \textbf{2016}, \emph{11},
  42--46\relax
\mciteBstWouldAddEndPuncttrue
\mciteSetBstMidEndSepPunct{\mcitedefaultmidpunct}
{\mcitedefaultendpunct}{\mcitedefaultseppunct}\relax
\EndOfBibitem
\bibitem[Gan \latin{et~al.}(2013)Gan, Shiue, Gao, Meric, Heinz, Shepard, Hone,
  Assefa, and Englund]{Chip-integrated-graphene}
Gan,~X.; Shiue,~R.-J.; Gao,~Y.; Meric,~I.; Heinz,~T.~F.; Shepard,~K.; Hone,~J.;
  Assefa,~S.; Englund,~D. {Chip-integrated ultrafast graphene photodetector
  with high responsivity}. \emph{Nature Photonics} \textbf{2013}, \emph{7},
  883--887\relax
\mciteBstWouldAddEndPuncttrue
\mciteSetBstMidEndSepPunct{\mcitedefaultmidpunct}
{\mcitedefaultendpunct}{\mcitedefaultseppunct}\relax
\EndOfBibitem
\bibitem[Ren \latin{et~al.}(2012)Ren, Zhang, Yao, Sun, Kaneko, Yan, Nanot, Jin,
  Kawayama, Tonouchi, Tour, and Kono]{Kono-tunable-properties}
Ren,~L.; Zhang,~Q.; Yao,~J.; Sun,~Z.; Kaneko,~R.; Yan,~Z.; Nanot,~S.; Jin,~Z.;
  Kawayama,~I.; Tonouchi,~M.; Tour,~J.~M.; Kono,~J. {Terahertz and infrared
  spectroscopy of gated large-area graphene}. \emph{Nano Letters}
  \textbf{2012}, \emph{12}, 3711--3715\relax
\mciteBstWouldAddEndPuncttrue
\mciteSetBstMidEndSepPunct{\mcitedefaultmidpunct}
{\mcitedefaultendpunct}{\mcitedefaultseppunct}\relax
\EndOfBibitem
\bibitem[Liu \latin{et~al.}(2014)Liu, Tian, Schwartz, Tok, Ren, and
  Bao]{Liu2014}
Liu,~N.; Tian,~H.; Schwartz,~G.; Tok,~J. B.-H.; Ren,~T.-L.; Bao,~Z. Large-Area,
  Transparent, and Flexible Infrared Photodetector Fabricated Using P-N
  Junctions Formed by N-Doping Chemical Vapor Deposition Grown Graphene.
  \emph{Nano Letters} \textbf{2014}, \emph{14}, 3702--3708\relax
\mciteBstWouldAddEndPuncttrue
\mciteSetBstMidEndSepPunct{\mcitedefaultmidpunct}
{\mcitedefaultendpunct}{\mcitedefaultseppunct}\relax
\EndOfBibitem
\bibitem[Wang \latin{et~al.}(2018)Wang, Zhang, Chen, Guo, Feng, Niu, Liu, Li,
  Lai, Sun, Liao, Wang, Chu, Ding, Xie, Di, and Wang]{Graphene_Lateral_PNJ}
Wang,~G. \latin{et~al.}  {Seamless lateral graphene p–n junctions formed by
  selective in situ doping for high-performance photodetectors}. \emph{Nature
  Communications} \textbf{2018}, \emph{9}, 5168\relax
\mciteBstWouldAddEndPuncttrue
\mciteSetBstMidEndSepPunct{\mcitedefaultmidpunct}
{\mcitedefaultendpunct}{\mcitedefaultseppunct}\relax
\EndOfBibitem
\bibitem[Castilla \latin{et~al.}(2019)Castilla, Terr{\'{e}}s, Autore, Viti, Li,
  Nikitin, Vangelidis, Watanabe, Taniguchi, Lidorikis, Vitiello, Hillenbrand,
  Tielrooij, and Koppens]{Castilla2019}
Castilla,~S.; Terr{\'{e}}s,~B.; Autore,~M.; Viti,~L.; Li,~J.; Nikitin,~A.~Y.;
  Vangelidis,~I.; Watanabe,~K.; Taniguchi,~T.; Lidorikis,~E.; Vitiello,~M.~S.;
  Hillenbrand,~R.; Tielrooij,~K.~J.; Koppens,~F.~H. {Fast and Sensitive
  Terahertz Detection Using an Antenna-Integrated Graphene pn Junction}.
  \emph{Nano Letters} \textbf{2019}, \emph{19}, 2765--2773\relax
\mciteBstWouldAddEndPuncttrue
\mciteSetBstMidEndSepPunct{\mcitedefaultmidpunct}
{\mcitedefaultendpunct}{\mcitedefaultseppunct}\relax
\EndOfBibitem
\bibitem[Mylnikov \latin{et~al.}(2023)Mylnikov, Titova, Kashchenko, Safonov,
  Zhukov, Semkin, Novoselov, Bandurin, and Svintsov]{Mylnikov2023}
Mylnikov,~D.~A.; Titova,~E.~I.; Kashchenko,~M.~A.; Safonov,~I.~V.;
  Zhukov,~S.~S.; Semkin,~V.~A.; Novoselov,~K.~S.; Bandurin,~D.~A.;
  Svintsov,~D.~A. {Terahertz Photoconductivity in Bilayer Graphene Transistors:
  Evidence for Tunneling at Gate-Induced Junctions}. \emph{Nano Letters}
  \textbf{2023}, \emph{23}, 220--226\relax
\mciteBstWouldAddEndPuncttrue
\mciteSetBstMidEndSepPunct{\mcitedefaultmidpunct}
{\mcitedefaultendpunct}{\mcitedefaultseppunct}\relax
\EndOfBibitem
\bibitem[Mueller \latin{et~al.}(2010)Mueller, Xia, and Avouris]{Mueller2010}
Mueller,~T.; Xia,~F.; Avouris,~P. {Graphene photodetectors for high-speed
  optical communications}. \emph{Nature Photonics} \textbf{2010}, \emph{4},
  297--301\relax
\mciteBstWouldAddEndPuncttrue
\mciteSetBstMidEndSepPunct{\mcitedefaultmidpunct}
{\mcitedefaultendpunct}{\mcitedefaultseppunct}\relax
\EndOfBibitem
\bibitem[Echtermeyer \latin{et~al.}(2014)Echtermeyer, Nene, Trushin, Gorbachev,
  Eiden, Milana, Sun, Schliemann, Lidorikis, Novoselov, and
  Ferrari]{Echtermeyer2014}
Echtermeyer,~T.~J.; Nene,~P.~S.; Trushin,~M.; Gorbachev,~R.~V.; Eiden,~A.~L.;
  Milana,~S.; Sun,~Z.; Schliemann,~J.; Lidorikis,~E.; Novoselov,~K.~S.;
  Ferrari,~A.~C. {Photothermoelectric and Photoelectric Contributions to Light
  Detection in Metal–Graphene–Metal Photodetectors}. \emph{Nano Letters}
  \textbf{2014}, \emph{14}, 3733--3742\relax
\mciteBstWouldAddEndPuncttrue
\mciteSetBstMidEndSepPunct{\mcitedefaultmidpunct}
{\mcitedefaultendpunct}{\mcitedefaultseppunct}\relax
\EndOfBibitem
\bibitem[Tielrooij \latin{et~al.}(2015)Tielrooij, Massicotte, Piatkowski,
  Woessner, Ma, Jarillo-Herrero, van Hulst, and Koppens]{Tielrooij2015}
Tielrooij,~K.~J.; Massicotte,~M.; Piatkowski,~L.; Woessner,~A.; Ma,~Q.;
  Jarillo-Herrero,~P.; van Hulst,~N.~F.; Koppens,~F. H.~L. {Hot-carrier
  photocurrent effects at graphene–metal interfaces}. \emph{Journal of
  Physics: Condensed Matter} \textbf{2015}, \emph{27}, 164207\relax
\mciteBstWouldAddEndPuncttrue
\mciteSetBstMidEndSepPunct{\mcitedefaultmidpunct}
{\mcitedefaultendpunct}{\mcitedefaultseppunct}\relax
\EndOfBibitem
\bibitem[Cai \latin{et~al.}(2014)Cai, Sushkov, Suess, Jadidi, Jenkins, Nyakiti,
  Myers-Ward, Li, Yan, Gaskill, Murphy, Drew, and Fuhrer]{Cai2014}
Cai,~X.; Sushkov,~A.~B.; Suess,~R.~J.; Jadidi,~M.~M.; Jenkins,~G.~S.;
  Nyakiti,~L.~O.; Myers-Ward,~R.~L.; Li,~S.; Yan,~J.; Gaskill,~D.~K.;
  Murphy,~T.~E.; Drew,~H.~D.; Fuhrer,~M.~S. {Sensitive room-temperature
  terahertz detection via the photothermoelectric effect in graphene}.
  \emph{Nature Nanotechnology} \textbf{2014}, \emph{9}, 814--819\relax
\mciteBstWouldAddEndPuncttrue
\mciteSetBstMidEndSepPunct{\mcitedefaultmidpunct}
{\mcitedefaultendpunct}{\mcitedefaultseppunct}\relax
\EndOfBibitem
\bibitem[Bandurin \latin{et~al.}(2018)Bandurin, Gayduchenko, Cao, Moskotin,
  Principi, Grigorieva, Goltsman, Fedorov, and Svintsov]{Bandurin2018}
Bandurin,~D.~A.; Gayduchenko,~I.; Cao,~Y.; Moskotin,~M.; Principi,~A.;
  Grigorieva,~I.~V.; Goltsman,~G.; Fedorov,~G.; Svintsov,~D. {Dual origin of
  room temperature sub-terahertz photoresponse in graphene field effect
  transistors}. \emph{Applied Physics Letters} \textbf{2018}, \emph{112},
  141101\relax
\mciteBstWouldAddEndPuncttrue
\mciteSetBstMidEndSepPunct{\mcitedefaultmidpunct}
{\mcitedefaultendpunct}{\mcitedefaultseppunct}\relax
\EndOfBibitem
\bibitem[Gayduchenko \latin{et~al.}(2018)Gayduchenko, Fedorov, Moskotin,
  Yagodkin, Seliverstov, Goltsman, {Yu Kuntsevich}, Rybin, Obraztsova, Leiman,
  Shur, Otsuji, and Ryzhii]{Gayduchenko2018}
Gayduchenko,~I.~A.; Fedorov,~G.~E.; Moskotin,~M.~V.; Yagodkin,~D.~I.;
  Seliverstov,~S.~V.; Goltsman,~G.~N.; {Yu Kuntsevich},~A.; Rybin,~M.~G.;
  Obraztsova,~E.~D.; Leiman,~V.~G.; Shur,~M.~S.; Otsuji,~T.; Ryzhii,~V.~I.
  {Manifestation of plasmonic response in the detection of sub-terahertz
  radiation by graphene-based devices}. \emph{Nanotechnology} \textbf{2018},
  \emph{29}, 245204\relax
\mciteBstWouldAddEndPuncttrue
\mciteSetBstMidEndSepPunct{\mcitedefaultmidpunct}
{\mcitedefaultendpunct}{\mcitedefaultseppunct}\relax
\EndOfBibitem
\bibitem[Semkin \latin{et~al.}(2022)Semkin, Mylnikov, Titova, Zhukov, and
  Svintsov]{Semkin}
Semkin,~V.; Mylnikov,~D.; Titova,~E.; Zhukov,~S.; Svintsov,~D. Gate-controlled
  polarization-resolving mid-infrared detection at metal{\textendash}graphene
  junctions. \emph{Applied Physics Letters} \textbf{2022}, \emph{120},
  191107\relax
\mciteBstWouldAddEndPuncttrue
\mciteSetBstMidEndSepPunct{\mcitedefaultmidpunct}
{\mcitedefaultendpunct}{\mcitedefaultseppunct}\relax
\EndOfBibitem
\bibitem[Zhou \latin{et~al.}(2018)Zhou, Raju, Li, Chan, Chai, and
  Yang]{Zhou_source-drain_width}
Zhou,~C.; Raju,~S.; Li,~B.; Chan,~M.; Chai,~Y.; Yang,~C.~Y. Self-Driven
  Metal–Semiconductor–Metal {WSe}$_2$ Photodetector with Asymmetric Contact
  Geometries. \emph{Advanced Functional Materials} \textbf{2018}, \emph{28},
  1802954\relax
\mciteBstWouldAddEndPuncttrue
\mciteSetBstMidEndSepPunct{\mcitedefaultmidpunct}
{\mcitedefaultendpunct}{\mcitedefaultseppunct}\relax
\EndOfBibitem
\bibitem[Auton \latin{et~al.}(2017)Auton, But, Zhang, Hill, Coquillat, Consejo,
  Nouvel, Knap, Varani, Teppe, Torres, and Song]{Auton2017}
Auton,~G.; But,~D.~B.; Zhang,~J.; Hill,~E.; Coquillat,~D.; Consejo,~C.;
  Nouvel,~P.; Knap,~W.; Varani,~L.; Teppe,~F.; Torres,~J.; Song,~A. {Terahertz
  Detection and Imaging Using Graphene Ballistic Rectifiers}. \emph{Nano
  Letters} \textbf{2017}, \emph{17}, 7015--7020\relax
\mciteBstWouldAddEndPuncttrue
\mciteSetBstMidEndSepPunct{\mcitedefaultmidpunct}
{\mcitedefaultendpunct}{\mcitedefaultseppunct}\relax
\EndOfBibitem
\bibitem[Brownless \latin{et~al.}(2020)Brownless, Zhang, and
  Song]{Geometric_rectifiers}
Brownless,~J.; Zhang,~J.; Song,~A. Graphene ballistic rectifiers: Theory and
  geometry dependence. \emph{Carbon} \textbf{2020}, \emph{168}, 201--208\relax
\mciteBstWouldAddEndPuncttrue
\mciteSetBstMidEndSepPunct{\mcitedefaultmidpunct}
{\mcitedefaultendpunct}{\mcitedefaultseppunct}\relax
\EndOfBibitem
\bibitem[Chen \latin{et~al.}(2022)Chen, Zhang, Feng, Xie, Jian, Li, Guo, Zhu,
  Li, Dong, Cui, Shi, and Xu]{Chen_Nonuniform_channel}
Chen,~J.; Zhang,~Z.; Feng,~J.; Xie,~X.; Jian,~A.; Li,~Y.; Guo,~H.; Zhu,~Y.;
  Li,~Z.; Dong,~J.; Cui,~Q.; Shi,~Z.; Xu,~C. 2D InSe Self-Powered Schottky
  Photodetector with the Same Metal in Asymmetric Contacts. \emph{Advanced
  Materials Interfaces} \textbf{2022}, \emph{9}, 2200075\relax
\mciteBstWouldAddEndPuncttrue
\mciteSetBstMidEndSepPunct{\mcitedefaultmidpunct}
{\mcitedefaultendpunct}{\mcitedefaultseppunct}\relax
\EndOfBibitem
\bibitem[Wei \latin{et~al.}(2020)Wei, Li, Wang, Liao, Dong, Xu, Zhu, Ang, Qiu,
  and Lee]{Wei2020}
Wei,~J.; Li,~Y.; Wang,~L.; Liao,~W.; Dong,~B.; Xu,~C.; Zhu,~C.; Ang,~K.-W.;
  Qiu,~C.-W.; Lee,~C. Zero-bias mid-infrared graphene photodetectors with bulk
  photoresponse and calibration-free polarization detection. \emph{Nature
  Communications} \textbf{2020}, \emph{11}, 6404\relax
\mciteBstWouldAddEndPuncttrue
\mciteSetBstMidEndSepPunct{\mcitedefaultmidpunct}
{\mcitedefaultendpunct}{\mcitedefaultseppunct}\relax
\EndOfBibitem
\bibitem[Wei \latin{et~al.}(2021)Wei, Xu, Dong, Qiu, and Lee]{Wei2021}
Wei,~J.; Xu,~C.; Dong,~B.; Qiu,~C.-W.; Lee,~C. Mid-infrared semimetal
  polarization detectors with configurable polarity transition. \emph{Nature
  Photonics} \textbf{2021}, \emph{15}, 614--621\relax
\mciteBstWouldAddEndPuncttrue
\mciteSetBstMidEndSepPunct{\mcitedefaultmidpunct}
{\mcitedefaultendpunct}{\mcitedefaultseppunct}\relax
\EndOfBibitem
\bibitem[Popov \latin{et~al.}(2015)Popov, Fateev, Ivchenko, and
  Ganichev]{Popov_2015}
Popov,~V.~V.; Fateev,~D.~V.; Ivchenko,~E.~L.; Ganichev,~S.~D.
  Noncentrosymmetric plasmon modes and giant terahertz photocurrent in a
  two-dimensional plasmonic crystal. \emph{Phys. Rev. B} \textbf{2015},
  \emph{91}, 235436\relax
\mciteBstWouldAddEndPuncttrue
\mciteSetBstMidEndSepPunct{\mcitedefaultmidpunct}
{\mcitedefaultendpunct}{\mcitedefaultseppunct}\relax
\EndOfBibitem
\bibitem[Yahniuk \latin{et~al.}(2022)Yahniuk, Budkin, Kazakov, Otteneder,
  Ziegler, Weiss, Mikhailov, Dvoretskii, Wojciechowski, Bel'kov, Knap, and
  Ganichev]{Ganichev_Grating}
Yahniuk,~I.; Budkin,~G.~V.; Kazakov,~A.; Otteneder,~M.; Ziegler,~J.; Weiss,~D.;
  Mikhailov,~N.~N.; Dvoretskii,~S.~A.; Wojciechowski,~T.; Bel'kov,~V.~V.;
  Knap,~W.; Ganichev,~S.~D. Terahertz Ratchet Effect in Interdigitated HgTe
  Structures. \emph{Phys. Rev. Appl.} \textbf{2022}, \emph{18}, 054011\relax
\mciteBstWouldAddEndPuncttrue
\mciteSetBstMidEndSepPunct{\mcitedefaultmidpunct}
{\mcitedefaultendpunct}{\mcitedefaultseppunct}\relax
\EndOfBibitem
\bibitem[Nikulin \latin{et~al.}(2021)Nikulin, Mylnikov, Bandurin, and
  Svintsov]{Nikulin2021}
Nikulin,~E.; Mylnikov,~D.; Bandurin,~D.; Svintsov,~D. {Edge diffraction,
  plasmon launching, and universal absorption enhancement in two-dimensional
  junctions}. \emph{Physical Review B} \textbf{2021}, \emph{103}, 085306\relax
\mciteBstWouldAddEndPuncttrue
\mciteSetBstMidEndSepPunct{\mcitedefaultmidpunct}
{\mcitedefaultendpunct}{\mcitedefaultseppunct}\relax
\EndOfBibitem
\bibitem[Long \latin{et~al.}(2019)Long, Wang, Wang, Zhou, Xia, Luo, Huang,
  Zhang, Yan, Fan, Wu, Chen, Lu, and Hu]{Long2019}
Long,~M.; Wang,~Y.; Wang,~P.; Zhou,~X.; Xia,~H.; Luo,~C.; Huang,~S.; Zhang,~G.;
  Yan,~H.; Fan,~Z.; Wu,~X.; Chen,~X.; Lu,~W.; Hu,~W. Palladium Diselenide
  Long-Wavelength Infrared Photodetector with High Sensitivity and Stability.
  \emph{{ACS} Nano} \textbf{2019}, \relax
\mciteBstWouldAddEndPunctfalse
\mciteSetBstMidEndSepPunct{\mcitedefaultmidpunct}
{}{\mcitedefaultseppunct}\relax
\EndOfBibitem
\bibitem[Lemme \latin{et~al.}(2011)Lemme, Koppens, Falk, Rudner, Park, Levitov,
  and Marcus]{Lemme}
Lemme,~M.~C.; Koppens,~F. H.~L.; Falk,~A.~L.; Rudner,~M.~S.; Park,~H.;
  Levitov,~L.~S.; Marcus,~C.~M. Gate-Activated Photoresponse in a Graphene
  p–n Junction. \emph{Nano Letters} \textbf{2011}, \emph{11},
  4134--4137\relax
\mciteBstWouldAddEndPuncttrue
\mciteSetBstMidEndSepPunct{\mcitedefaultmidpunct}
{\mcitedefaultendpunct}{\mcitedefaultseppunct}\relax
\EndOfBibitem
\bibitem[Badioli \latin{et~al.}(2014)Badioli, Woessner, Tielrooij, Nanot,
  Navickaite, Stauber, {Garc{\'{i}}a De Abajo}, and Koppens]{Badioli2014}
Badioli,~M.; Woessner,~A.; Tielrooij,~K.~J.; Nanot,~S.; Navickaite,~G.;
  Stauber,~T.; {Garc{\'{i}}a De Abajo},~F.~J.; Koppens,~F.~H. {Phonon-mediated
  mid-infrared photoresponse of graphene}. \emph{Nano Letters} \textbf{2014},
  \emph{14}, 6374--6381\relax
\mciteBstWouldAddEndPuncttrue
\mciteSetBstMidEndSepPunct{\mcitedefaultmidpunct}
{\mcitedefaultendpunct}{\mcitedefaultseppunct}\relax
\EndOfBibitem
\bibitem[Nishiyama \latin{et~al.}(2022)Nishiyama, Nishimura, Nishioka, Ueno,
  Iwamoto, and Nagashio]{Nishiyama}
Nishiyama,~W.; Nishimura,~T.; Nishioka,~M.; Ueno,~K.; Iwamoto,~S.; Nagashio,~K.
  Is the Bandgap of Bulk {PdSe}$_2$ Located Truly in the Far-Infrared Region
  Determination by Fourier-Transform Photocurrent Spectroscopy. \emph{Advanced
  Photonics Research} \textbf{2022}, \emph{3}, 2200231\relax
\mciteBstWouldAddEndPuncttrue
\mciteSetBstMidEndSepPunct{\mcitedefaultmidpunct}
{\mcitedefaultendpunct}{\mcitedefaultseppunct}\relax
\EndOfBibitem
\bibitem[Lee \latin{et~al.}(2008)Lee, Balasubramanian, Weitz, Burghard, and
  Kern]{Lee_contacts_and_edges}
Lee,~E.~J.; Balasubramanian,~K.; Weitz,~R.~T.; Burghard,~M.; Kern,~K. {Contact
  and edge effects in graphene devices}. \emph{Nature Nanotechnology}
  \textbf{2008}, \emph{3}, 486--490\relax
\mciteBstWouldAddEndPuncttrue
\mciteSetBstMidEndSepPunct{\mcitedefaultmidpunct}
{\mcitedefaultendpunct}{\mcitedefaultseppunct}\relax
\EndOfBibitem
\bibitem[Afzal \latin{et~al.}(2020)Afzal, Dastgeer, Iqbal, Gautam, and
  Faisal]{Afzal_PdSe2_photovoltaic}
Afzal,~A.~M.; Dastgeer,~G.; Iqbal,~M.~Z.; Gautam,~P.; Faisal,~M.~M.
  High-Performance p-BP/n-PdSe2 Near-Infrared Photodiodes with a Fast and
  Gate-Tunable Photoresponse. \emph{ACS Applied Materials \& Interfaces}
  \textbf{2020}, \emph{12}, 19625--19634\relax
\mciteBstWouldAddEndPuncttrue
\mciteSetBstMidEndSepPunct{\mcitedefaultmidpunct}
{\mcitedefaultendpunct}{\mcitedefaultseppunct}\relax
\EndOfBibitem
\bibitem[Dai \latin{et~al.}(2020)Dai, Chen, Wang, Long, Shang, Hu, Li, Ge,
  Zhang, Zhai, Fu, and Hu]{Dai}
Dai,~M.; Chen,~H.; Wang,~F.; Long,~M.; Shang,~H.; Hu,~Y.; Li,~W.; Ge,~C.;
  Zhang,~J.; Zhai,~T.; Fu,~Y.; Hu,~P. Ultrafast and Sensitive Self-Powered
  Photodetector Featuring Self-Limited Depletion Region and Fully Depleted
  Channel with van der Waals Contacts. \emph{{ACS} Nano} \textbf{2020},
  \emph{14}, 9098--9106\relax
\mciteBstWouldAddEndPuncttrue
\mciteSetBstMidEndSepPunct{\mcitedefaultmidpunct}
{\mcitedefaultendpunct}{\mcitedefaultseppunct}\relax
\EndOfBibitem
\bibitem[Schulman \latin{et~al.}(2018)Schulman, Arnold, and
  Das]{Contact_engineering}
Schulman,~D.~S.; Arnold,~A.~J.; Das,~S. Contact engineering for 2D materials
  and devices. \emph{Chem. Soc. Rev.} \textbf{2018}, \emph{47},
  3037--3058\relax
\mciteBstWouldAddEndPuncttrue
\mciteSetBstMidEndSepPunct{\mcitedefaultmidpunct}
{\mcitedefaultendpunct}{\mcitedefaultseppunct}\relax
\EndOfBibitem
\bibitem[De~Nicola \latin{et~al.}(2018)De~Nicola, Puthiya~Purayil, Spirito,
  Miscuglio, Tantussi, Tomadin, De~Angelis, Polini, Krahne, and
  Pellegrini]{Polini_fractal}
De~Nicola,~F.; Puthiya~Purayil,~N.~S.; Spirito,~D.; Miscuglio,~M.;
  Tantussi,~F.; Tomadin,~A.; De~Angelis,~F.; Polini,~M.; Krahne,~R.;
  Pellegrini,~V. Multiband Plasmonic Sierpinski Carpet Fractal Antennas.
  \emph{ACS Photonics} \textbf{2018}, \emph{5}, 2418--2425\relax
\mciteBstWouldAddEndPuncttrue
\mciteSetBstMidEndSepPunct{\mcitedefaultmidpunct}
{\mcitedefaultendpunct}{\mcitedefaultseppunct}\relax
\EndOfBibitem
\bibitem[Campbell \latin{et~al.}(2019)Campbell, Sell, Jenkins, Whiting, Fan,
  and Werner]{Campbell_MLDesign}
Campbell,~S.~D.; Sell,~D.; Jenkins,~R.~P.; Whiting,~E.~B.; Fan,~J.~A.;
  Werner,~D.~H. {Review of numerical optimization techniques for meta-device
  design [Invited]}. \emph{Optical Materials Express} \textbf{2019}, \emph{9},
  1842\relax
\mciteBstWouldAddEndPuncttrue
\mciteSetBstMidEndSepPunct{\mcitedefaultmidpunct}
{\mcitedefaultendpunct}{\mcitedefaultseppunct}\relax
\EndOfBibitem
\bibitem[Ermolaev \latin{et~al.}(2021)Ermolaev, Voronin, Tatmyshevskiy,
  Mazitov, Slavich, Yakubovsky, Tselin, Mironov, Romanov, Markeev, Kruglov,
  Novikov, Vyshnevyy, Arsenin, and Volkov]{Ermolaev2021}
Ermolaev,~G.~A.; Voronin,~K.~V.; Tatmyshevskiy,~M.~K.; Mazitov,~A.~B.;
  Slavich,~A.~S.; Yakubovsky,~D.~I.; Tselin,~A.~P.; Mironov,~M.~S.;
  Romanov,~R.~I.; Markeev,~A.~M.; Kruglov,~I.~A.; Novikov,~S.~M.;
  Vyshnevyy,~A.~A.; Arsenin,~A.~V.; Volkov,~V.~S. {Broadband Optical Properties
  of Atomically Thin PtS2 and PtSe2}. \emph{Nanomaterials} \textbf{2021},
  \emph{11}, 3269\relax
\mciteBstWouldAddEndPuncttrue
\mciteSetBstMidEndSepPunct{\mcitedefaultmidpunct}
{\mcitedefaultendpunct}{\mcitedefaultseppunct}\relax
\EndOfBibitem
\bibitem[Kischkat \latin{et~al.}(2012)Kischkat, Peters, Gruska, Semtsiv,
  Chashnikova, Klinkm\"{u}ller, Fedosenko, Machulik, Aleksandrova,
  Monastyrskyi, Flores, and Masselink]{Kischkat}
Kischkat,~J.; Peters,~S.; Gruska,~B.; Semtsiv,~M.; Chashnikova,~M.;
  Klinkm\"{u}ller,~M.; Fedosenko,~O.; Machulik,~S.; Aleksandrova,~A.;
  Monastyrskyi,~G.; Flores,~Y.; Masselink,~W.~T. Mid-infrared optical
  properties of thin films of aluminum oxide, titanium dioxide, silicon
  dioxide, aluminum nitride, and silicon nitride. \emph{Appl. Opt.}
  \textbf{2012}, \emph{51}, 6789--6798\relax
\mciteBstWouldAddEndPuncttrue
\mciteSetBstMidEndSepPunct{\mcitedefaultmidpunct}
{\mcitedefaultendpunct}{\mcitedefaultseppunct}\relax
\EndOfBibitem
\end{mcitethebibliography}

\renewcommand{\theequation} {S\arabic{equation}}
\renewcommand{\thefigure} {S\arabic{figure}}
\setcounter{figure}{0}
\newpage
\section{Supporting Information}
\section{I. Devices fabrication and characterization}

\begin{figure}[ht]
    \includegraphics[width=0.5\textwidth]{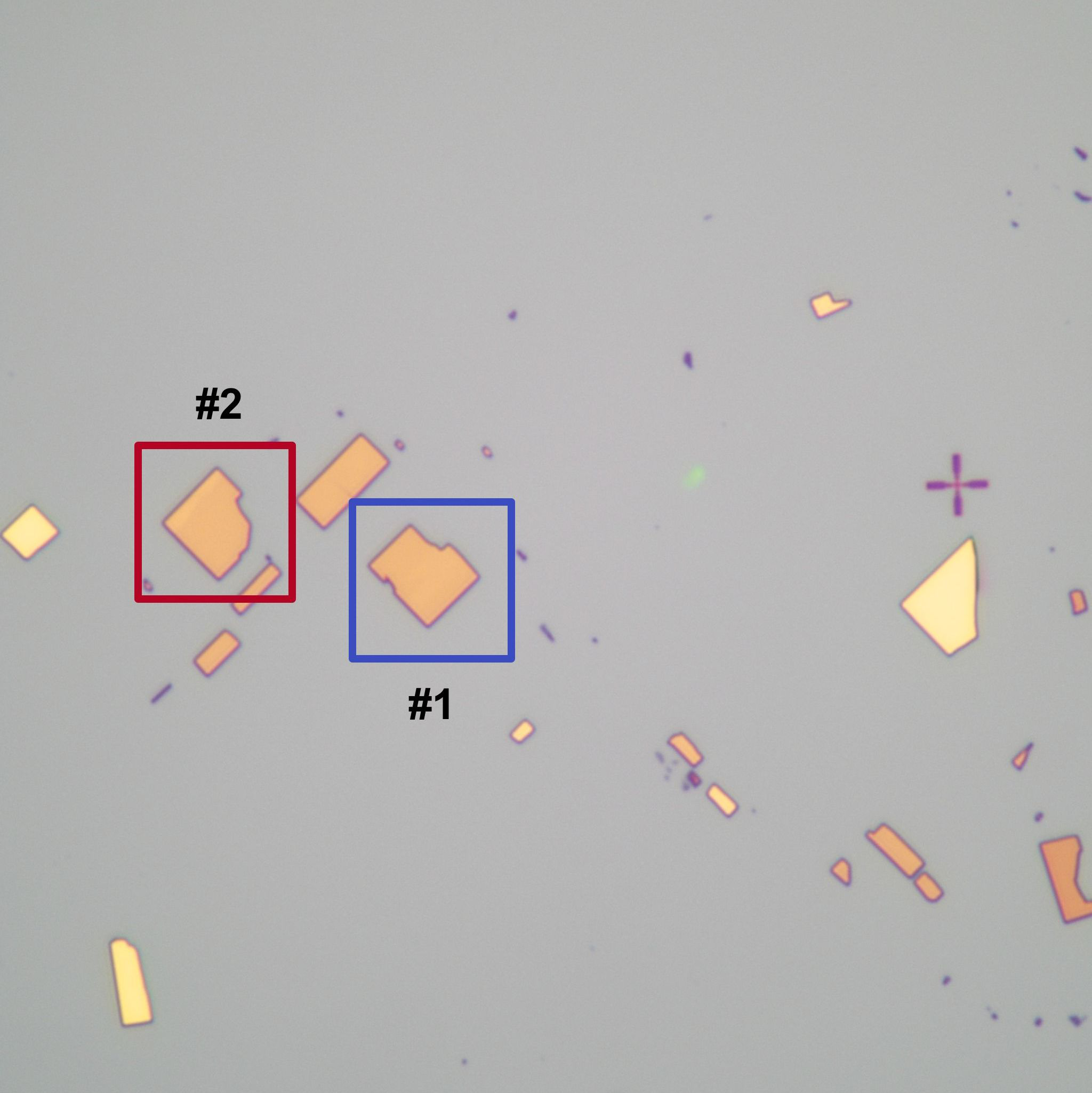}
    \caption{\label{figS1}Nearby 2d PdSe$_2$ flakes of similar sizes and optically equal thickness, which were used to fabricate detectors. Flakes coated with PMMA on a substrate before lithography.}
\end{figure}

\begin{figure}[ht]
    \includegraphics[width=0.7\linewidth]{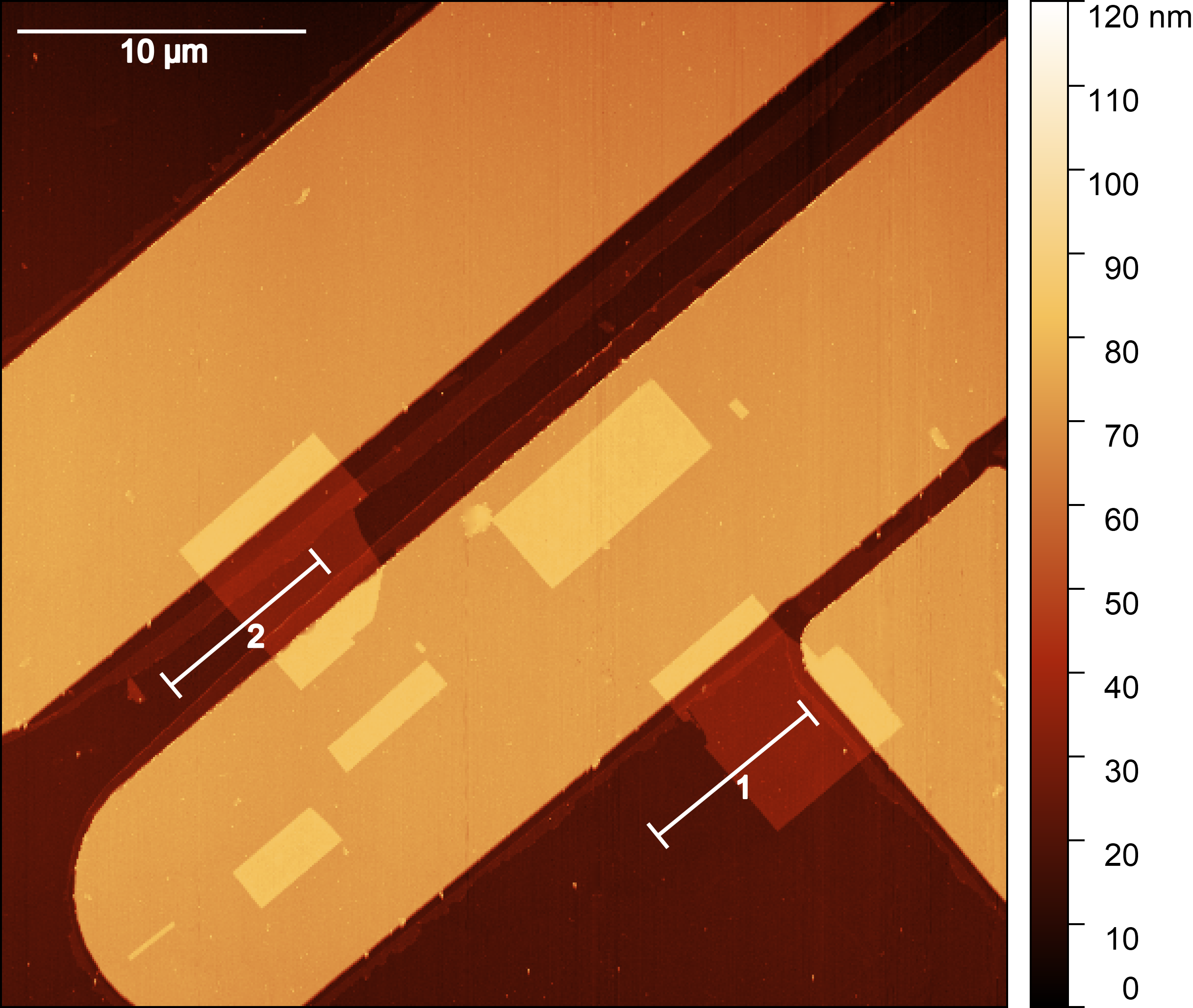}
    \includegraphics[width=0.8\linewidth]{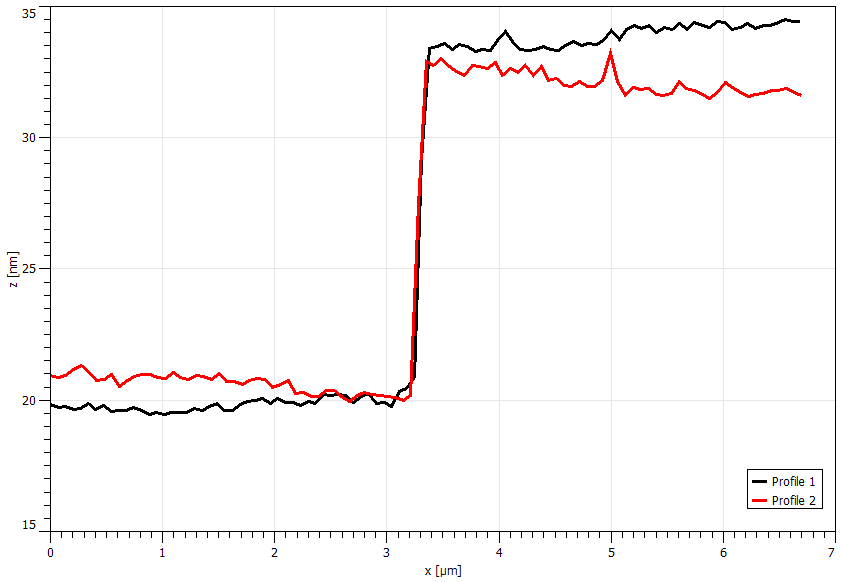}
    \caption{\label{figS2}AFM scan of detectors (top) and PdSe$_2$ flakes profiles (bottom).}
\end{figure}

\clearpage
\section{II. Photocurrent measurement details}

\begin{figure}[ht]
    \includegraphics[width=\textwidth]{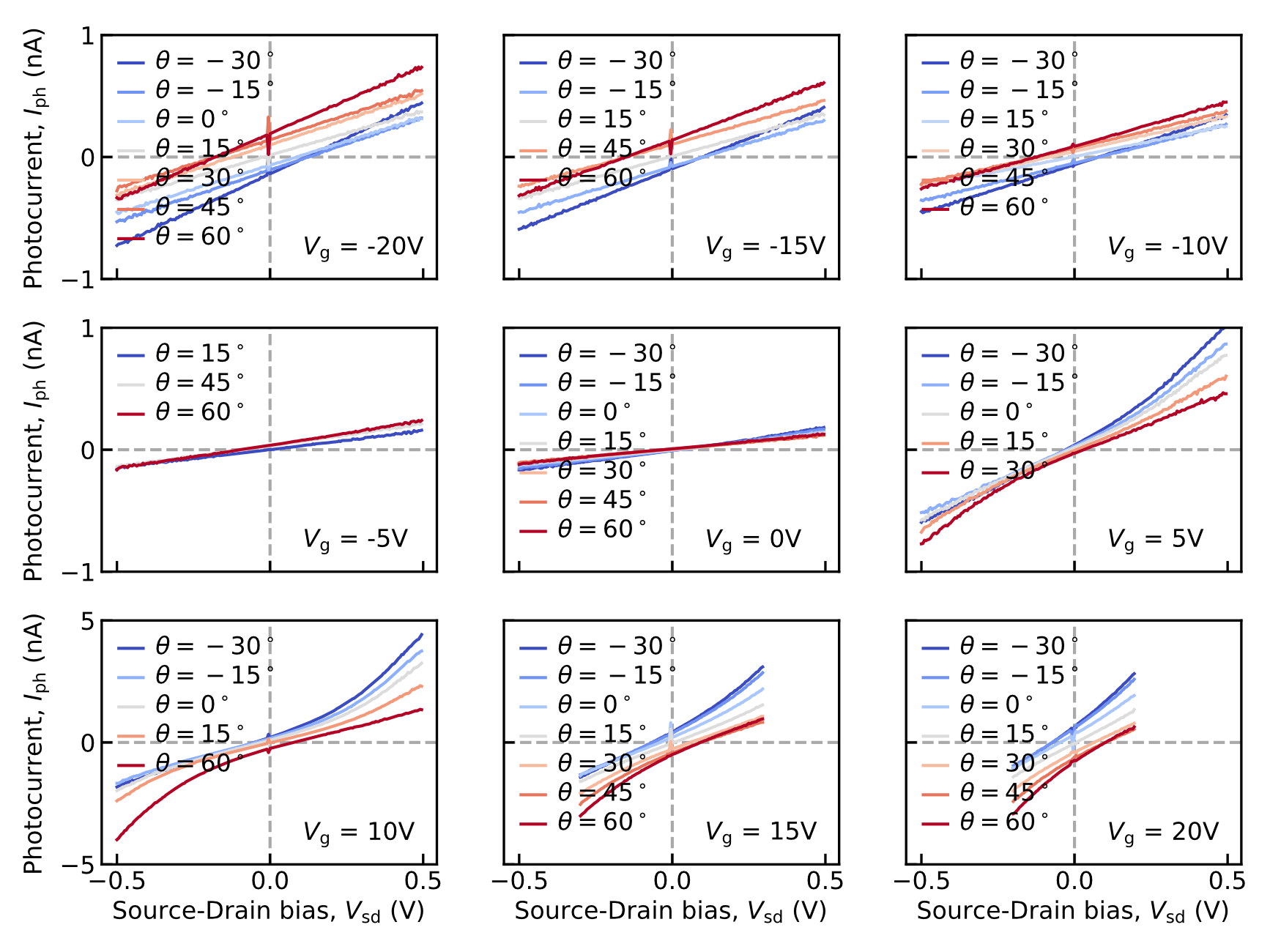}
    \caption{\label{figS3}Series of photocurrent $I_{\rm ph}$ dependencies on the bias voltage $V_{\rm sd}$ at different gate voltages $V_{\rm g}$ and polarization angles $\theta$ for corner-type detector $\#1$.}
\end{figure}

\begin{figure}[ht]
    \includegraphics[width=\textwidth]{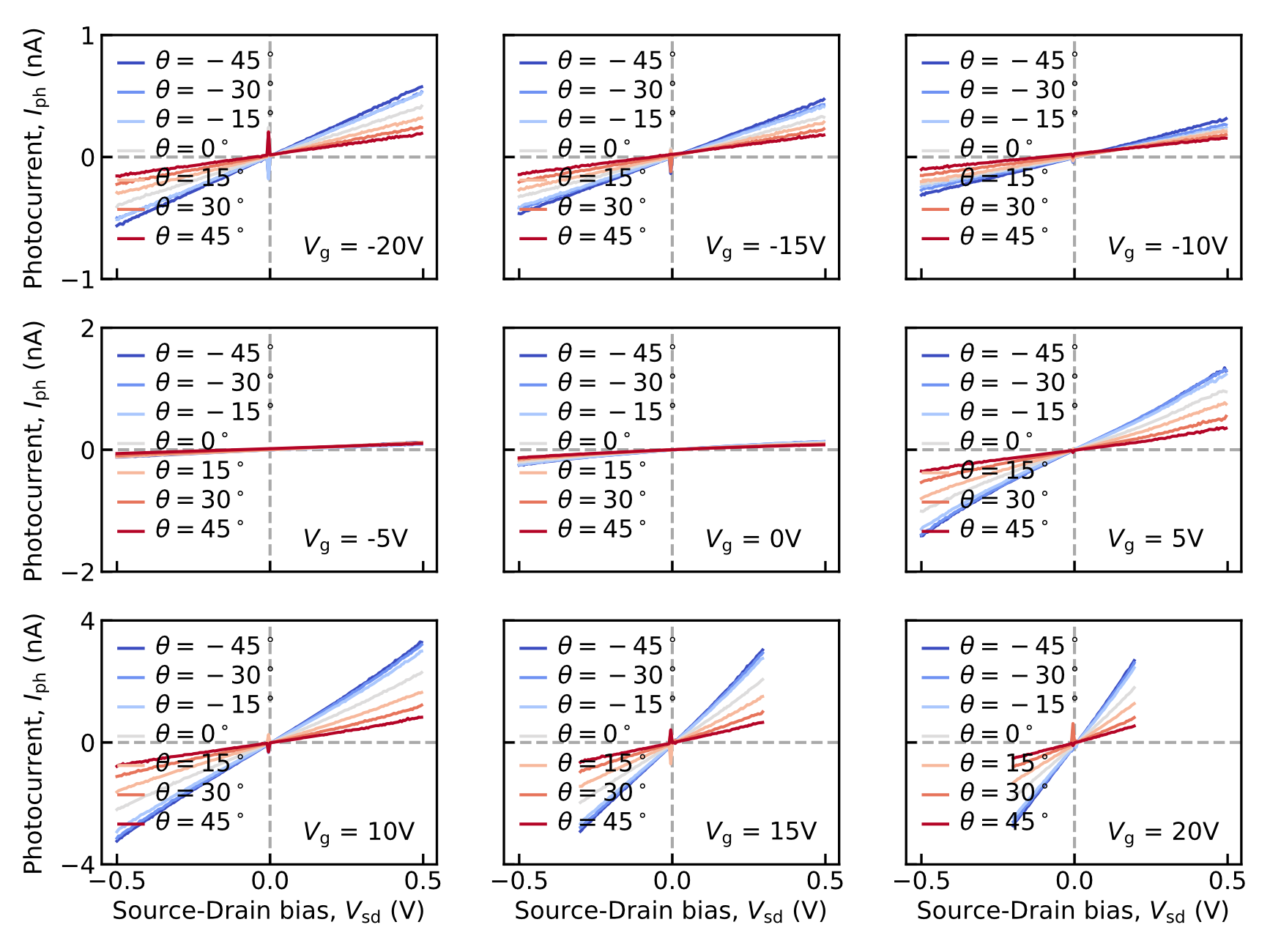}
    \caption{\label{figS4}Series of photocurrent $I_{\rm ph}$ dependencies on the bias voltage $V_{\rm sd}$ at different gate voltages $V_{\rm g}$ and polarization angles $\theta$ for reference slit-type detector $\#2$.}
\end{figure}

\begin{figure}[ht]
    \includegraphics[width=\textwidth]{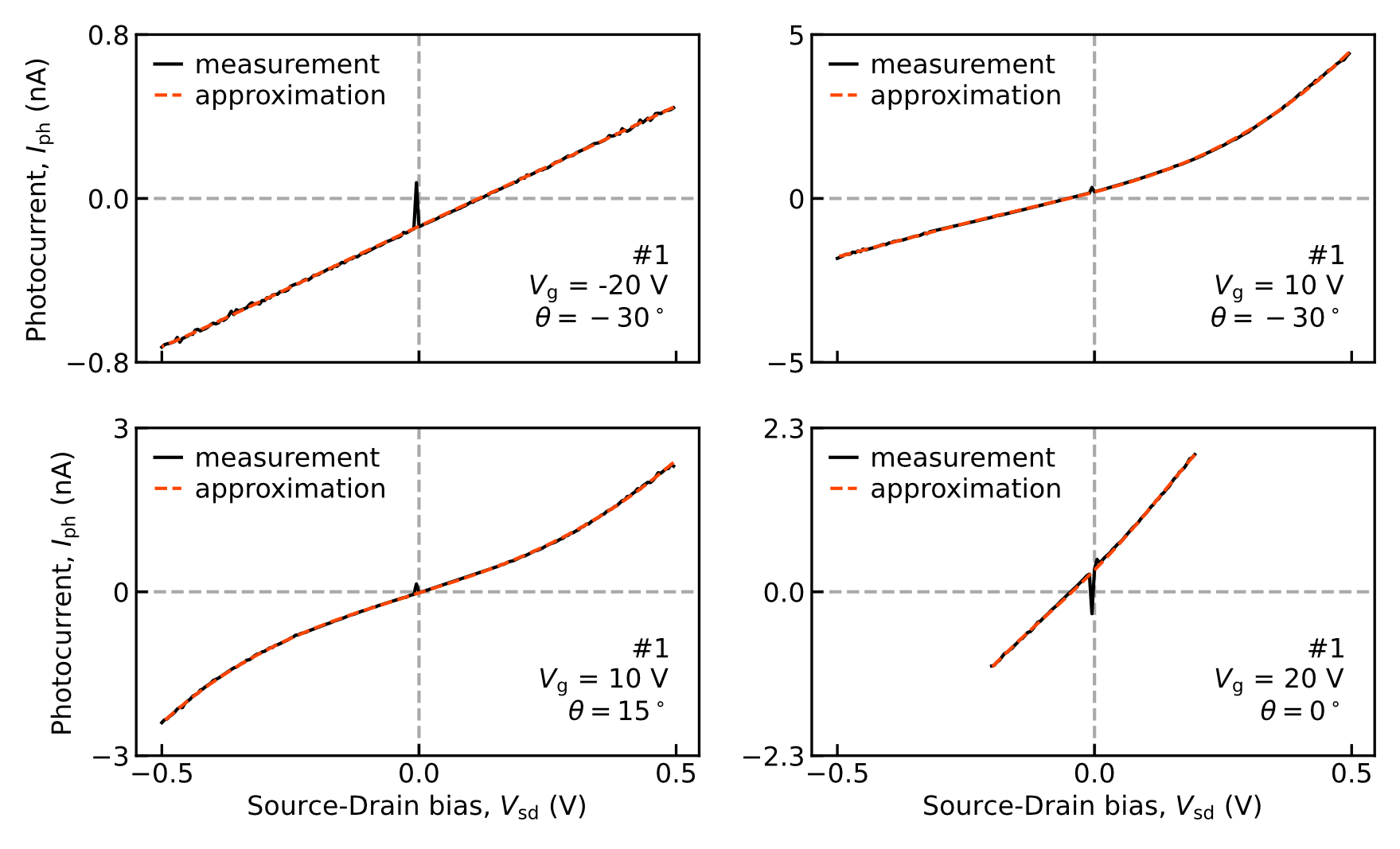}
    \caption{\label{figS5}Several examples of experimental data approximation for the photocurrent $I_{\rm ph}$ dependencies on the bias voltage $V_{\rm sd}$. The spikes near zero are clearly visible, because of which the approximation had to be performed. It was manually verified that all other dependencies are approximated as well as those presented here.}
\end{figure}

\begin{figure}[ht]
    \includegraphics[width=\textwidth]{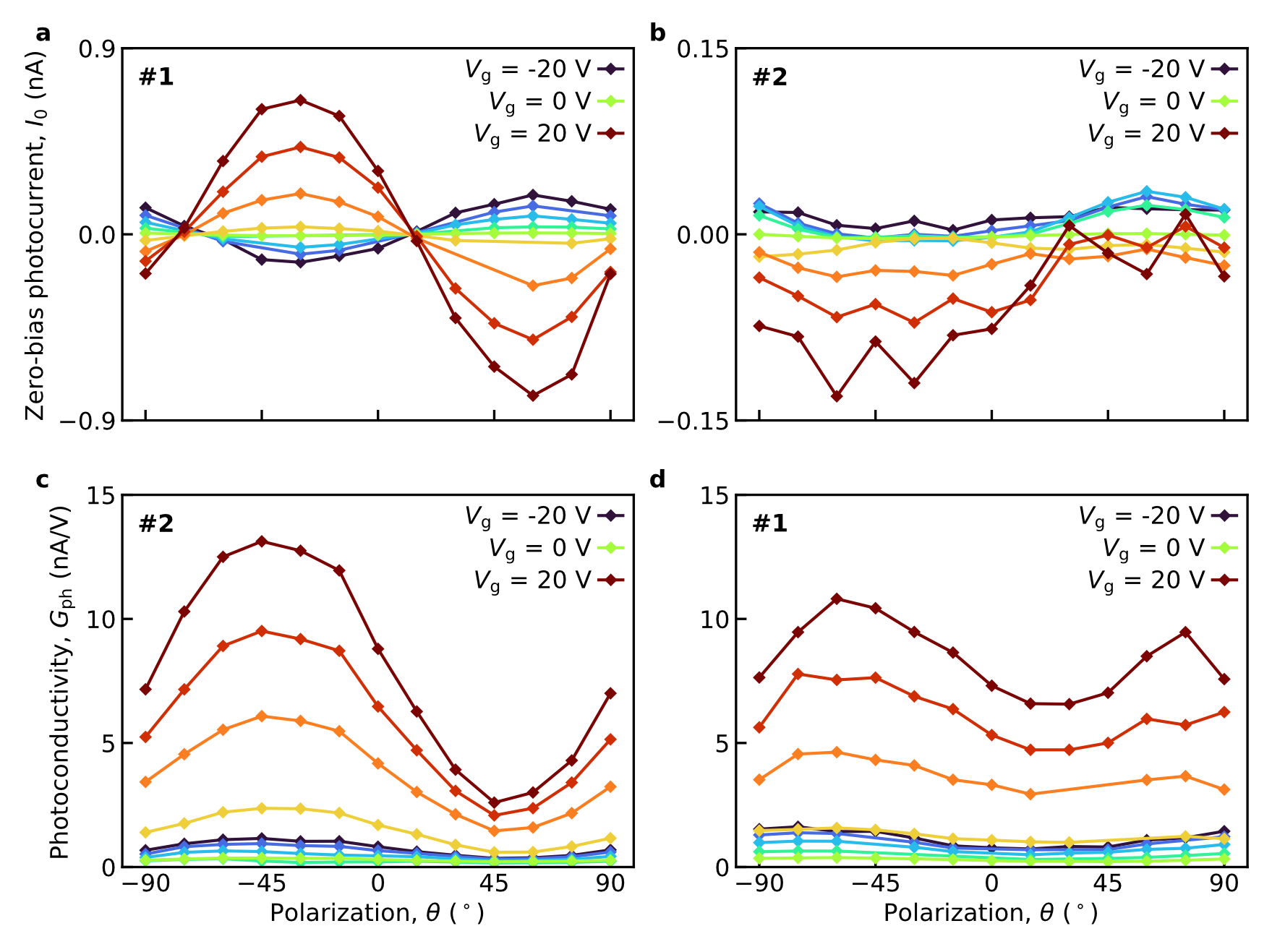}
    \caption{\label{figS6}{\bf a,b,} Extracted from $I_{\rm ph}( V_{\rm sd})$ series zero-bias photocurrent $I_0$ for $\#1$ ({\bf a}, Fig. 2b in the main text) and for $\#2$ ({\bf b}) at different gate voltages $V_{\rm g}$ with step $\Delta V_{\rm g}~=~5~$V. {\bf c,d,} Extracted from $I_{\rm ph}( V_{\rm sd})$ series photoconductivity $G_{\rm ph}$ for $\#2$ ({\bf c}, Fig. 2d in the main text) and for $\#1$ ({\bf d}) at different $V_{\rm g}$ with step $\Delta V_{\rm g}~=~5~$V.}
\end{figure}

\clearpage
\section{III. Simulation details}

\begin{figure}[ht]
    \includegraphics[width=\textwidth]{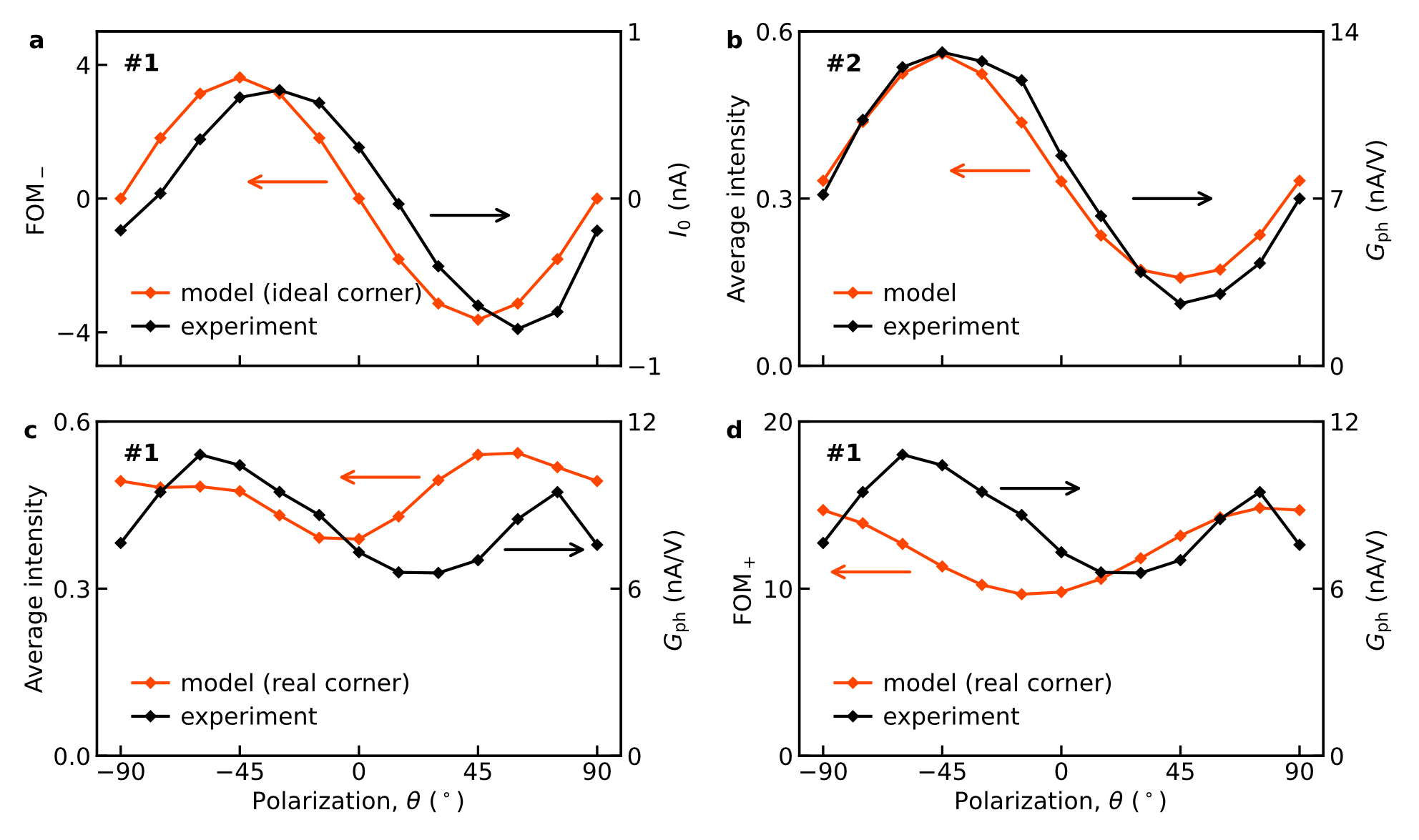}
    \caption{\label{figS7}{\bf a} Comparison of the measured zero-bias photocurrent $I_0(\theta)$ with the simulation of the normalized difference of average local intensities at source and drain FOM$_-$ for ideal square corner-type detector without a gap between the contacts. {\bf b, c} Comparison of the measured photoconductivity $G_{\rm ph}(\theta)$ with the simulation of total average intensity at PdSe$_2$ flake for the device \#2 ({\bf b}) and for the device \#1 ({\bf c}). {\bf d} Comparison of the measured photoconductivity $G_{\rm ph}(\theta)$ with the simulation of normalized sum of average local intensities at source and drain FOM$_+$ for the device \#1. All experimental curves were measured at gate voltage $V_{\rm g}$ = 20 V.}
\end{figure}

\section{IV. Optimization of corner-type detector}
\begin{figure}[ht]
    \includegraphics[width=\textwidth]{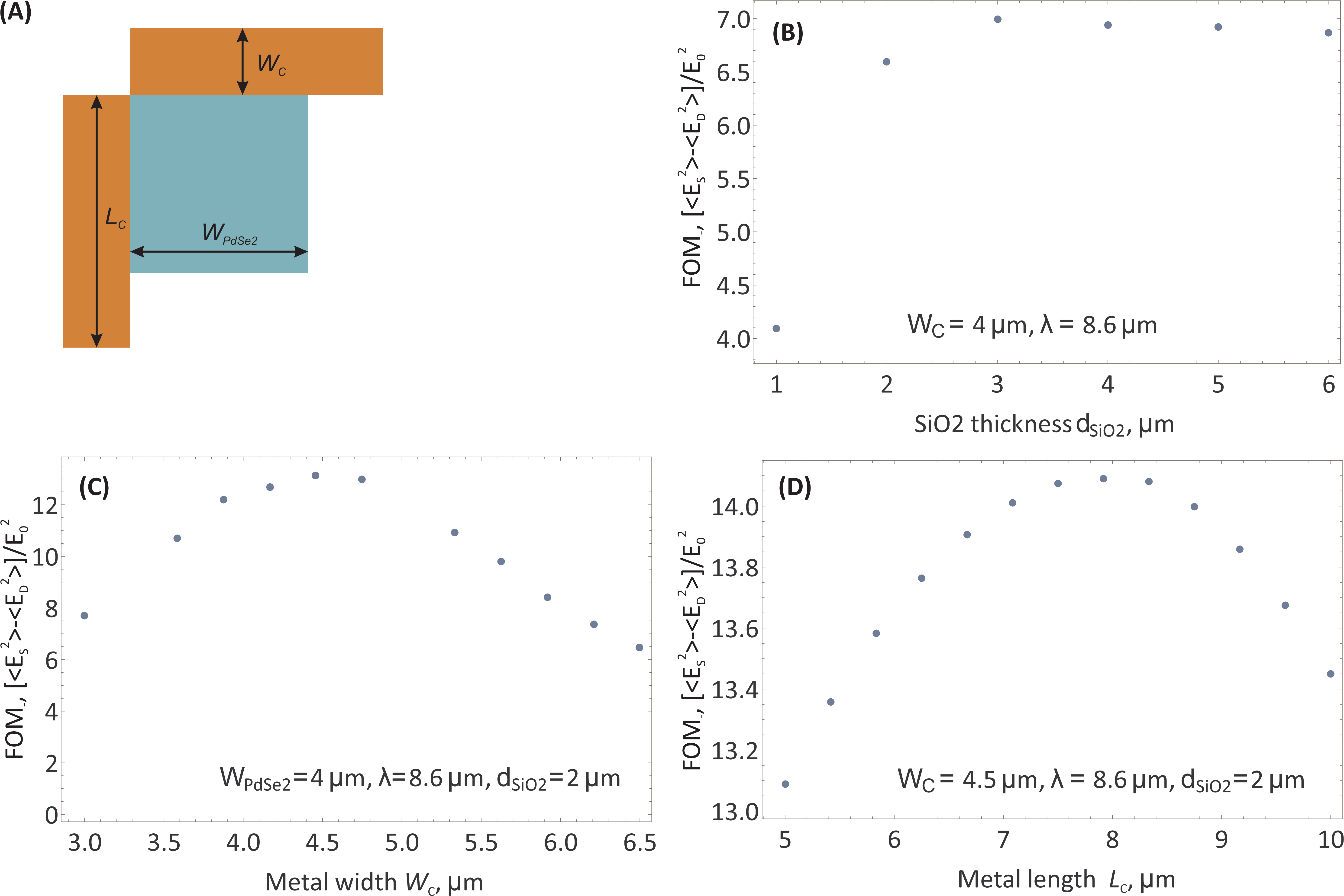}
    \caption{\label{figS8}{\bf Simulation optimization of corner-type detector} {\bf (a)} Scheme of the structure considered in simulation optimization {\bf (c-d)} Dependences of FOM$_-$ on various structure parameters: SiO$_2$ thickness (b), metal pad width $W_C$ (c), metal pad length $L_C$ (d).}
\end{figure}

In this section, we present the results on simulation optimization for the corner-type photodetector. We aim to maximize FOM$_-$ being the difference of local intensities at source and drain contacts, normalized by the incident intensity. In our simulations, we vary the thickness of SiO$_2$ substrate, the width of metal contacts $W_C$ and their length $L_C$. The size of PdSe$_2$ flake is held fixed and equal to $W_{PdSe2} = 4$ $\mu$m. The scheme of the structure used in our simulation is shown in Fig.~\ref{figS8}. We intentionally consider a perfect photodetector being symmetric with respect to the diagonal of its channel, and ignore the details of real experimental structure such as non-equal sides of the channel and inter-contact gap.

It appears that the most crucial parameter for optimization is the thickness of SiO$_2$ substrate [Fig.~\ref{figS8} (b)]. If the oxide thickness is well below the radiation wavelength, FOM$_-$ takes very low values due to destructive interference of the incident wave and the wave reflected from high-index Si substrate. At $d_{SiO2}\sim 2$ $\mu$m, the figure of merit almost saturates at its maximum value. No interference pattern is observed upon variation of oxide thickness due to large phonon absorption at our particular wavelength $\lambda=8.6$ $\mu$m. 

Fixing the technologically achievable SiO$_2$ thickness to 2 $\mu$m, we vary the contact width $W_C$ in Fig.~\ref{figS8} (c). The maximum FOM$_-\approx 12$ is achieved at $W_C\approx4.5$ $\mu$m. This width corresponds to the lowest dipole antenna-type resonance at metal width. Further optimization with respect to metal length $L_C$ [Fig.~\ref{figS8} (d)] raises the FOM$_-$ up to $\sim 14$. The optimum is now achieved at $L_C \approx 8$ $\mu$m. Such length corresponds to antenna-type anti-resonance in the contact parallel to the polarization of the incident radiation.

To conclude, an optimized detector should reside on a thick oxide layer to avoid the interference-type suppression of the incident field. The width of metal contacts should be tuned approximately to $\lambda/2$, while their length -- approximately to $\lambda$. This provides maximum field enhancement at the 'active' Schottky contact and maximum field suppression at the 'passive' one.

\end{document}